\documentclass[prb,showpacs,twocolumn,amsmath,amssymb,floatfix,superscriptaddress]{revtex4-1}

\usepackage{multirow}
\usepackage{array}
\usepackage{graphicx}
\usepackage{bm}
\usepackage[usenames]{color}
\usepackage{dsfont}
\usepackage[utf8]{inputenc}

\newcommand{\beq}{\begin{equation}}
\newcommand{\eeq}{\end{equation}}
\newcommand{\beqa}{\begin{eqnarray}}
\newcommand{\eeqa}{\end{eqnarray}}

\newcommand{\edit}[1] {\textcolor{red}{#1}}  

\begin{document}

\title{Strain-induced bound states in transition-metal dichalcogenide bubbles}

\author{L. Chirolli} 
\email{luca.chirolli@gmail.com}
\affiliation{IMDEA Nanoscience, C/Faraday 9, ES-28049 Madrid, Spain}
\author{E. Prada}
\affiliation{Departamento de F\'isica de la Materia Condensada, Condensed Matter Physics Center (IFIMAC) and 
Instituto Nicol\'as Cabrera, Universidad Aut\'onoma de Madrid, E-28049 Madrid, Spain}
\author{F. Guinea} 
\affiliation{IMDEA Nanoscience, C/Faraday 9, ES-28049 Madrid, Spain}
\affiliation{School of Physics and Astronomy, University of Manchester, Manchester, M13 9PY, United Kingdom}
\author{R. Rold\'an}
\author{P. San-Jose}
\affiliation{Materials Science Factory, Instituto de Ciencia de Materiales de Madrid (ICMM-CSIC), Sor Juana In\'es de la Cruz 3, 28049 Madrid, Spain}

\date{\today}

\begin{abstract}
We theoretically study the formation of single-particle bound states confined by strain at the center of bubbles in monolayers of transition-metal 
dichalcogenides (TMDs). Bubbles ubiquitously form in two-dimensional crystals on top of a substrate by the competition between van der Waals 
forces and the hydrostatic pressure exerted by trapped fluid. This leads to strong strain at the center of the bubble that reduces the bangap locally, 
creating potential wells for the electrons that confine states inside. We simulate the spectrum versus the bubble radius for the four semiconducting 
group VI \edit{TMDs}, MoS$_2$, WSe$_2$, WS$_2$ and MoSe$_2$, and find an overall Fock-Darwin spectrum of bubble bound states, characterised 
by small deviations compatible with Berry curvature effects. We analyse the density of states, the state degeneracies, orbital structure and optical 
transition rules. Our results show that elastic bubbles in these materials are remarkably efficient at confining photocarriers.
\end{abstract}

\maketitle 

{\it Introduction.---} The possibility of applying mechanical stress to a system is an extremely useful tool to locally tune its electronic and optical 
properties. Atomically-thin two-dimensional (2D) crystals and van der Waals (vdW) materials~\cite{NG05,Novoselov2016,RoldanRev2017} are particularly 
suited for strain engineering and straintronics~\cite{Levy2010} applications. Their 2D surface can be easily accessed by indenters and probes, they can be 
integrated with nanopatterned substrates, and stacked and/or twisted with respect to other 2D crystals, developing tensions between them that depend on 
the lattice mismatches, twist angles or pressure ~\cite{Alden:PNAS13,San-Jose:PRB14,Yankowitz:NC16}. Moreover, the mechanical properties can be 
typically described with the elastic theory of membranes and strongly differ from those of their 3D counterparts. Together with graphene and phosphorene, 
group-VI transition metal dichalcogenides (\edit{TMDs}) in particular have been shown, both in theory\cite{RG15,Rostami_PRB_2015} and in experiments 
\cite{CS13,He2013,Plechinger2015}, to be ideal candidates for strain engineering. One of the peculiar properties of \edit{TMDs} under strain is the reduction 
of the band gap\cite{Lloyd2016} in the regions where tensile strain is applied. This property can be exploited to funnel excitons created upon irradiation to 
well defined regions characterised by a reduced gap\cite{FL12, San-Jose_PRX_2016}. Upon exciton recombination,  quantised single-photon emission is 
achieved\cite{Kumar2015,PB2016,Branny2016,Branny2017,Shepard2017,Ye2017,Chaste2018,Tripathi2018}, that represents a key ingredient for quantum 
information processing and quantum technologies \cite{OBrien2009}. At the basis of single-photon quantized emission is the localisation of bound states by 
strain-induced quantum confinement, that can be also used as spin-valley qubits\cite{Liu2014,YueWu2016,Szechenyi2018}. 

In this work we theoretically investigate the formation of bound states in small bubbles that naturally arise in monolayer TMDs deposited over a substrate.
It has been theoretically and experimentally shown that substances, typically hydrocarbons, become trapped between the crystal and the atomically flat substrate 
and lead, through the competition between vdW forces and elastic energy, to the formation of bubbles with shape and internal pressure that are universal for 2D 
crystals\cite{Khestanova_NC_2016}. Typical sizes range from sub-10 nm to sub-micron, with very high internal pressure of the order of GPa. This leads to very 
strong and highly localised elastic strains. By simulating bubbles of up to 10 nm in size we show here that bubble strain leads to a sizeable suppression of the local 
bandgap in the bubble, that acts as a potential well for conduction band carriers. As a result, strongly localised, valley-degenerate electronic states quickly develop 
in the bubbles. The spectrum of confined states is non-trivial, and inherits features of the low energy structure of the conduction band. A similar strain-induced 
confinement effect has been recently explored in which the strain is extrinsic, produced in the TMD by indentation with nanopillars patterned on the 
substrate~\cite{Li_NC_2015,PB2016,Branny2016,Branny2017,Chaste2018,Tripathi2018,Burkard_2018}.

Our findings can be generalised to bubbles with non-universal profiles, or to interacting photocarriers relevant for the problem of exciton funnelling. More 
generally, they provide a basis for studies on quantum confinement of spin-valley qubits, and may contribute to a characterisation of novel optical selection 
rules in 2D systems\cite{Louie2018,Xiao2018}.

\begin{figure}[t]
\begin{center}
\includegraphics[width=\columnwidth]{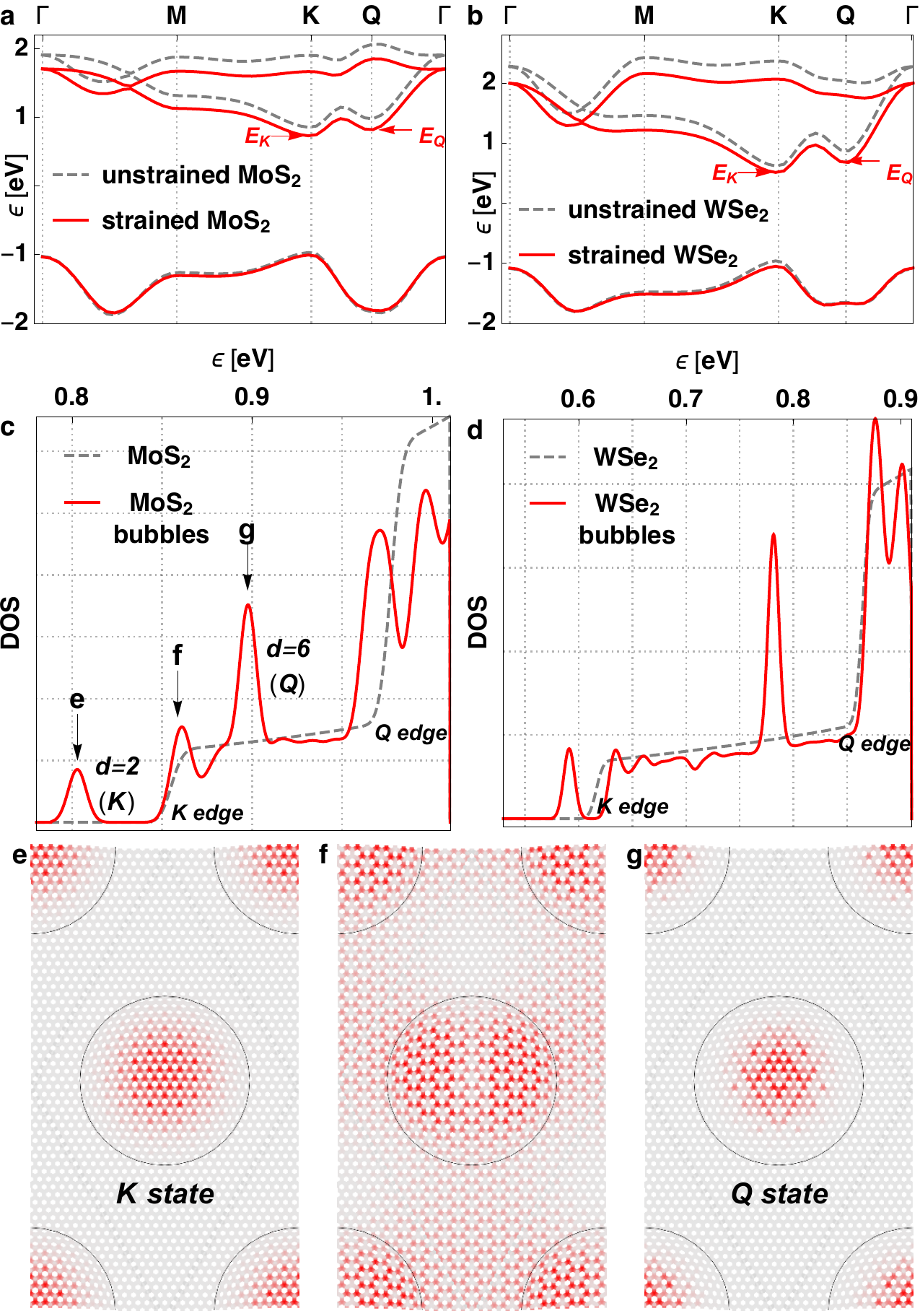}
\caption{(color online)  Bands in MoS$_2$ (a) and WSe$_2$ (b) without strains (dashed) and under uniform biaxial strain $\epsilon_\mathrm{max}$ (solid red), 
see Eq.~\eqref{emax}  (spin-orbit interaction [SOI] not included).  (c) Density of states (DOS) integrated over all space for an MoS$_2$ monolayer without 
strains (dashed) and in the presence of a superlattice of bubbles (solid red), with bubble radius $R=3$nm, height-over-radius aspect ratio $h_{\rm max}/R=0.14$, 
and separation between bubbles of $L=9.5$nm. With the bubbles, DOS peaks appear inside the gap close to the conduction band edge at $0.85$ eV (with 
semiconducting gap $E_g=1.82$eV). (d) The same as (a) for a WSe$_2$ monolayer. Here the conduction band edge is at $0.62$eV and $E_g=1.57$eV. (e-g) 
Local density of states plotted on the MoS$_2$ lattice with the periodic superlattice of small bubbles (black circles) for the three energies marked in (c). SOI is not 
considered in the present results. \label{Fig1}}
\end{center}
\end{figure}

{\it Elasticity theory and bubble profile.---} 
The universal shape of bubbles of atomically thin crystals has been obtained self-consistently in Ref.~[\onlinecite{Khestanova_NC_2016}] assuming constant 
volume, with the crystal profile adapting to the competition between the hydrostatic pressure of the trapped fluid and the vdW attraction of the 2D crystal 
and the substrate at the edges. In-plane stiffness and the energy associated to out-of-plane bending determine the rigidity of the 2D crystal. In-plane stiffness is 
described by the theory of elasticity\cite{Landau} through Lam\'e coefficients, $\lambda$ and $\mu$, or alternatively through the Young's modulus, $Y$, and the 
Poisson's ratio, $\nu$. The out-of-plane bending is accounted for by the bending rigidity $\kappa$. For curvature radii beyond a scale $\ell_{\rm anh}\sim 
\sqrt{Y/\kappa}$ (on the order of 1nm for MoS$_2$), bending rigidity can be neglected.\cite{Khestanova_NC_2016} This is the case of all realistic bubbles 
observed in experiments. The total energy is then reduced to the contributions from the in-plane elastic energy and the energy of the trapped fluid. 
The bubble profile is then obtained by energy minimisation. (An alternative but equivalent approach considers a constant pressure $P$ inside the bubble, 
with the bubble profile resulting from the condition of static equilibrium)

Assuming circular bubbles of radius $R$ an isotropic solution can be described in terms of the height profile 
\begin{equation}
h(r)=h_{\rm max}\tilde{h}(r/R),
\end{equation}
with $\tilde{h}(r)$ a normalised dimensionless function. Using dimensional arguments the total energy is written as
\begin{equation}
E_{\rm tot} = E_{\rm el} - PV = c_1 Y \frac{h^4_{\rm max}}{R^2} - c_V PR^2 h_{\rm max},
\end{equation}
where $c_1$ and $c_V$ are dimensionless constants that describe the elastic energy and the volume, respectively, associated to the bubble 
shape defined by the function $\tilde{h}(r)$. Minimising the energy with respect to $h_{\rm max}$ we obtain for the aspect ratio of the bubble
\begin{equation}\label{Eq:AspectRatio}
\frac{h_{\rm max}}{R}=\left(\frac{c_VPR}{4c_1Y}\right)^{1/3},
\end{equation}
that gives $E_{\rm tot}[\tilde{h}]=-\frac{3}{4}\left[c_V^4[\tilde{h}]/c_1[\tilde{h}]\right]^{1/3}\left(P^4R^{10}/(4Y)\right)^{1/2}$. The profile $\tilde{h}$
is obtained by minimisation of the total energy and its value turns out to be independent of $R$, $P$, and $Y$ [\onlinecite{Khestanova_NC_2016}]. 
The associated in-plane displacement, $u_r(r)$ can be obtained by solving a Poisson equation with the source given in terms of the function $h(r)$ [\onlinecite{Khestanova_NC_2016}]. The self-consistent solution is accurately approximated by a parabolic profile
\begin{equation}
h(r)\approx h_{\rm max}(1-(r/R)^2),
\end{equation}
which in turn yields a radial displacement of the form
\begin{equation}
u_r\approx\frac{h^2_{\rm max}}{2R}\frac{\lambda+3\mu}{\lambda+2\mu}\frac{r}{R}\left(1-\frac{r^2}{R^2}\right).
\end{equation}
We use this approximate profile to simulate the bubble electronic structure through a tight-binding model. Note that the maximum biaxial strain 
$\epsilon_\mathrm{max}$ is reached at the center of the bubble, and that it only depends on the aspect ratio, not the radius,
\begin{equation}
\epsilon_\mathrm{max} = \frac{1}{2}\frac{\lambda+3\mu}{\lambda+2\mu}\left(\frac{h_{\rm max}}{R}\right)^2.
\label{emax}
\end{equation}

{\it Tight-binding model.---} 
Most monolayer TMDs, such as MoS${}_2$ or WSe${}_2$,  are direct bandgap semiconductors, with a gap of the order of 2 eV placed at the two inequivalent $K$ and $K'$ points of the Brillouin 
zone (BZ) and characterised by peculiar optical selection rules, that promote circular dichroism\cite{Mak2012,Zeng2012}. Their conduction band also 
presents a second minimum placed at the so-called $Q$ point of the BZ, along the $\Gamma$-$K$ direction, which is six-fold degenerate (see Fig. 
\ref{Fig1}[a,b]). Contrary to what happens in other 2D materials like graphene or black phosphorus, the conduction and valence bands of TMDs present 
a very rich orbital contribution.\cite{CG13,Silva_AS_2016} They are made by hybridisation of the $d$ orbitals of the metal (Mo or W), and the $p$ orbitals 
of the chalcogen (S or Se). In addition, TMDs are characterised by strong spin-orbit interaction (SOI), which leads to a large splitting of the valence band 
at the $K$ and $K'$ points of the BZ, as well as at the conduction band edge at the $Q$ point.\cite{Silva_AS_2016}

The electronic band structure of TMDs is well described by a Slater-Koster tight-binding approximation throughout the BZ.\cite{CG13} The model 
includes 11 orbitals: the five $d$ orbitals of the metal atom and  the six $p$ orbitals of the two chalcogen atoms in the unit cell. We 
refer the reader to Refs. [\onlinecite{CG13,Silva_AS_2016}] for details of the model. The tight-binding Hamiltonian can be expressed in real space as
\begin{eqnarray}
\label{Eq:H}
H
&=&
\sum_{i,\mu\nu}
\epsilon_{\mu,\nu} c^\dagger_{i,\mu}c_{i,\nu}
+
\sum_{ ij,\mu\nu}
{[t_{ij,\mu\nu} c^\dagger_{i,\mu}c_{j,\nu}+{\rm H.c.}]},
\end{eqnarray} 
where $c^\dagger_{i,\mu}(c_{i,\mu})$ creates (annihilates) an electron in the unit cell $i$
in the atomic orbital $\mu=1,\ldots,11$. 

The above tight-binding approximation is especially useful to study modifications in the band structure due to realistic deformations of the crystal. 
Strain leads to variations in the interatomic bond lengths, which modifies the corresponding hopping terms $t_{ij,\mu\nu}({\bf r}_{ij})$ as
\begin{eqnarray}\label{Eq:hopping}
t_{ij,\mu\nu}({\bf r}_{ij})&=&t_{ij,\mu\nu}({\bf r}_{ij}^0)
\exp 
\left[-\Lambda_{ij,\mu\nu}
\left(
\frac{|{\bf r}_{ij}|}{|{\bf r}_{ij}^0|}-1
\right)
\right].
\end{eqnarray}
Here $|{\bf r}_{ij}^0|$ is the separation between atoms labelled by $(i,\mu)$ and $(j,\nu)$ in the absence of strain,
$|{\bf r}_{ij}|$ the distance in the presence of strain, and $\Lambda_{ij,\mu\nu}=-d\ln t_{ij,\mu\nu}(r)/d\ln(r)|_{r=|{\bf r}_{ij}^0|}$ 
is the bond-resolved local electron-phonon coupling. The intersite vectors in the strained lattice ${\bf r}_{ij}$ are obtained from 
the strain tensor $\bm \varepsilon$ as ${\bf r}_{ij}={\bf r}_{ij}^0+{\bm \varepsilon}\cdot {\bf r}_{ij}^0$. Following previous works,
\cite{Rostami_PRB_2015,San-Jose_PRX_2016} we make use the Wills-Harrison argument~\cite{Harrison_1999} that predicts 
a dependence $\Lambda_{ij,\mu\nu} \approx l_\mu+l_\nu+1$, where $l_\mu$($l_\nu$) is the absolute value of the angular 
momentum of the orbital $\mu$ ($\nu$), leading to bond-dependent electron-phonon couplings. The most prominent effect of 
tensile strain in TMDs is a suppression of their gap. Red lines in Figure \ref{Fig1}(a,b) illustrate the effect. Under realistic biaxial 
strains $\epsilon\lesssim 5-10\%$ it is accurate to describe the variation of the gap $\Delta E_\mathrm{gap}$ by a proportionality 
constant $\gamma>0$,
\begin{equation}
\Delta E_\mathrm{gap} = -\gamma\,\epsilon.
\label{gamma}
\end{equation} 
For MoS$_2$, $\gamma\approx 6.4$eV, or 64meV per 1\% of strain. 
\begin{figure}[t]
\begin{center}
\includegraphics[width=\columnwidth]{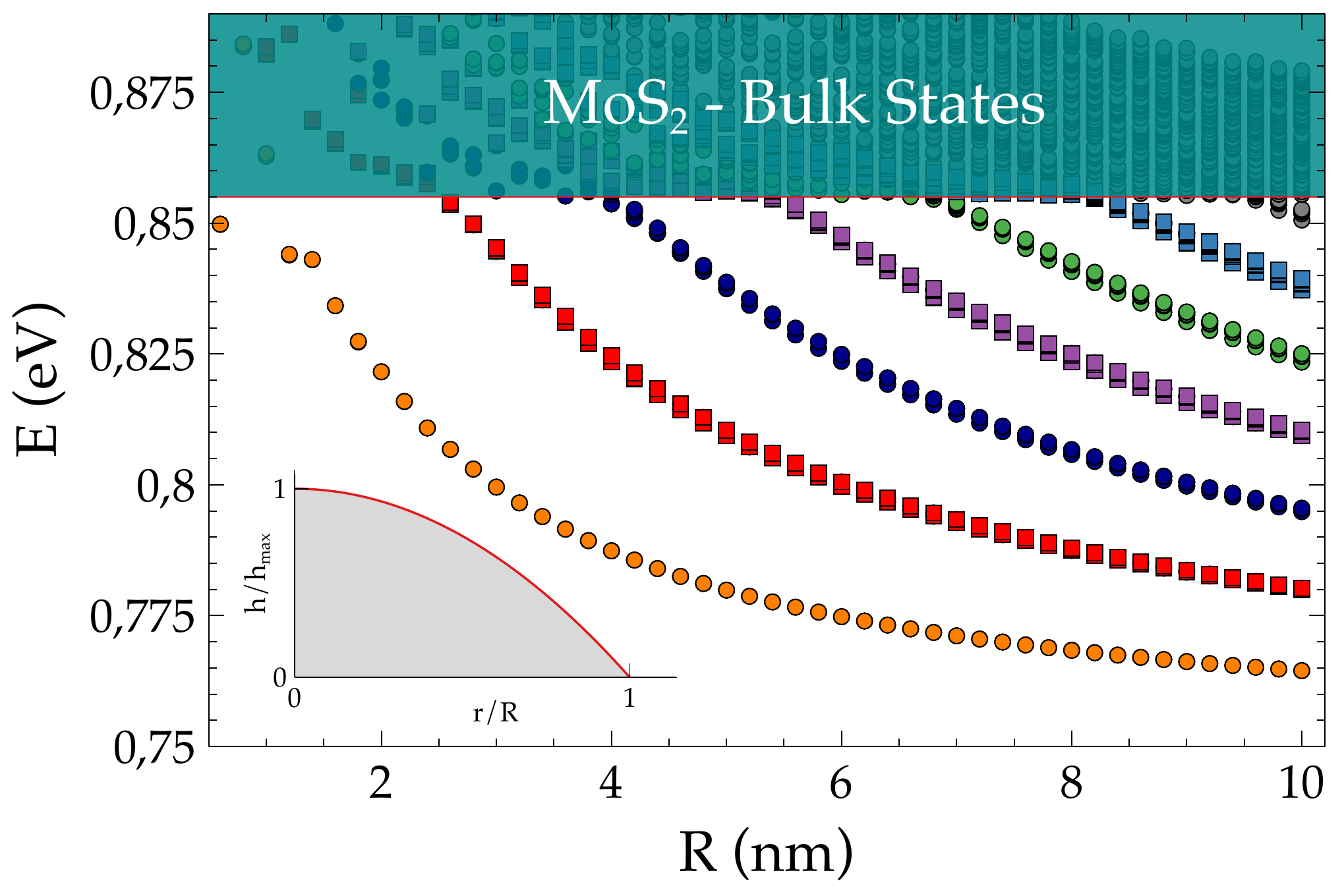}
\caption{Evolution of the energy of the subgap bound states inside a monolayer MoS$_2$ bubble as a function of bubble radius $R$. 
The aspect ratio is fixed to $h_{\rm max}/R=0.14$. The spectrum is compatible with an overall Fock-Darwin spectrum, with alternating 
even (circle marker) and odd (square marker) parity states. States independently detach from the $K$ and $K'$ valleys, giving an additional 
double degeneracy to each level. The overall degeneracy $d=2,4,6,\ldots$ (ignoring spin) is represented with a color code: orange for 
$d=2$, red for the $d=4$, blue for $d=6$, etc.. Inset: sketch of the bubble profile $h(r)=h_{\rm max}(1-(r/R)^2)$. 
\label{Fig2}}
\end{center}
\end{figure}

{\it Spectrum of a circular bubble.---} 
We now consider circular bubbles of radius $R$ described using the above formalism. We construct a periodic triangular superlattice of them, with a period $L$ such that both $L/R$ and $h_{\rm max}/R$ are fixed. We simulate the electronic structure of the system using the MathQ package \cite{MathQ}. Spin and spin-orbit interaction are neglected at this stage. In Fig. \ref{Fig1} we show the bands and the corresponding density of states (DOS) for unstrained MoS$_2$ and WSe$_2$ (dashed lines in panels a-d). Note in panels (c,d) the two DOS steps (`K edge' and `Q edge') corresponding to the K/K' and Q band minima. We also show the bands under uniform strain $\epsilon_\mathrm{max}$ representative of the bubble center (solid red in a-b), and the global density of states in the presence of a bubble superlattice of a small $R=3$ nm radius (red lines in panels c-d). We assume a bubble aspect ratio $h_{\rm max}/R =0.14$ similar to those observed in experiments\cite{Khestanova_NC_2016}, and a superlattice period $L/R\approx 3.1$. 

Notably, even for such small bubbles, we find that in both materials a quasi-flat band of localised states, two per bubble, drops into the uniform gap from the conduction band  (see `e' arrow). These are valley degenerate states localised inside the bubbles that detach from the valence $K$, $K'$ points by virtue of the strain-induced suppression of the local gap. Panel e shows the probability density of one such valley degenerate state (black circles mark the bubble contours). Panel f corresponds to a state at a higher energy, not energetically detached from the band continuum, and therefore delocalised outside the bubbles. We also show in panel g a state from the strong peak visible between the K and Q edges (`g' arrow in panel c). This state is rather remarkable. While it lives within a delocalised background of states, it is strongly localised inside the bubbles. It is in fact six-fold degenerate, $d=6$, in contrast to the $d=2$ degeneracy of state e. Its origin is completely analogous to the latter, however, but instead of detaching from the $K,K'$ points, the six states per bubble detach from the Q-edge formed by the six $Q$ points in the bandstructure. Given the smoothness of the bubble, the Q states remain mostly orthogonal to the delocalised K,K' in their background, which allow them to remain highly localised.

To better understand the formation of localised states we compute the low-energy spectrum of the bubble superlattice at the $\Gamma$ point around the conduction edge as the bubble radius $R$ increases. We fix the bubble aspect ratio to $L/R\approx 6.2$ to keep the bubbles well separated. The results are shown in Fig. \ref{Fig2}. We find a dense continuum of states above the gap, plus a set of non-dispersive discrete levels that again detach from the K edge into the bulk gap for any $R\gtrsim 3$nm (their value is essentially insensitive to $L$ as soon as $L/R\gtrsim 2$). The Q states are not readily visible in this plot, as they lie inside the bulk continuum.

\begin{figure}[t]
\begin{center}
\includegraphics[width=\columnwidth]{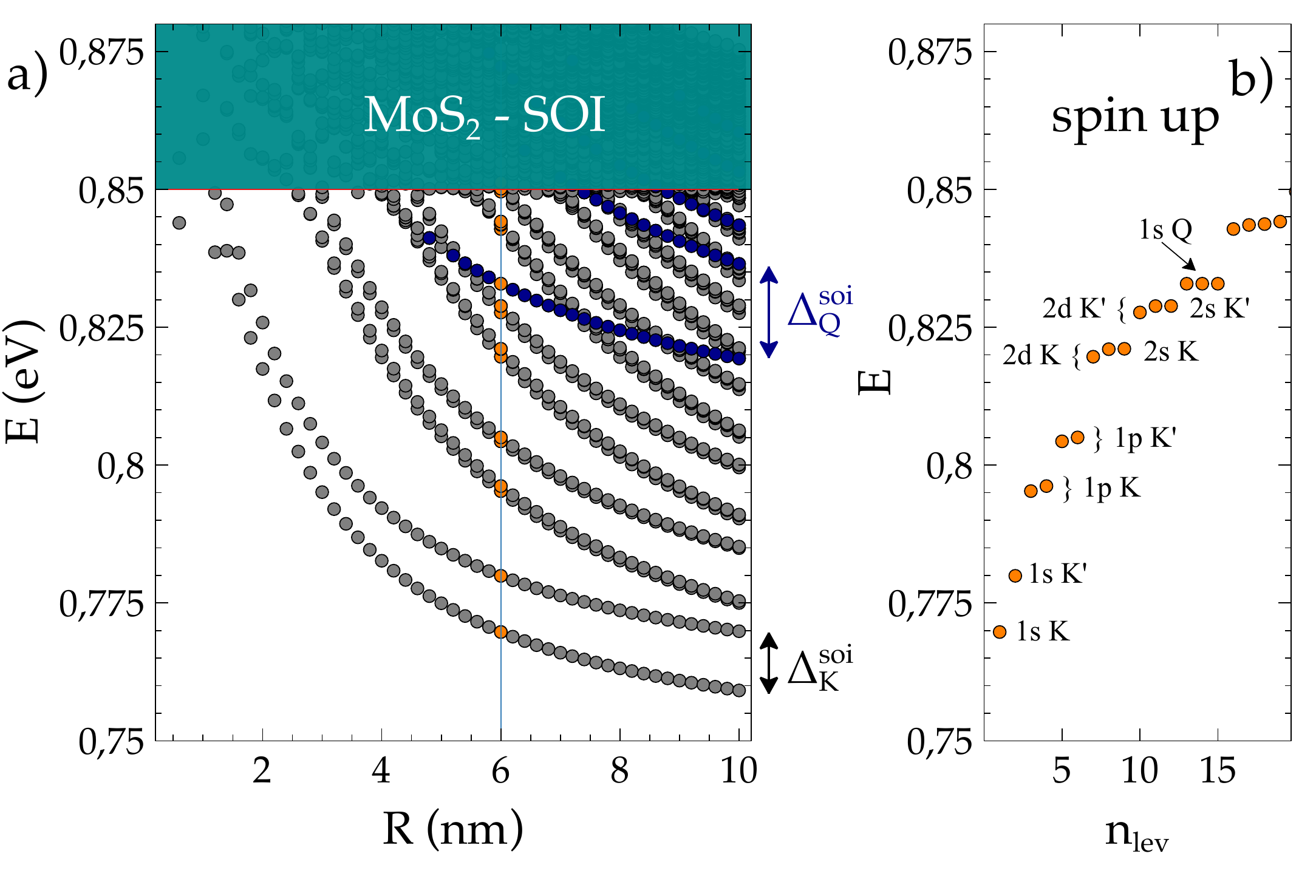}
\caption{Effect of the spin-orbit interaction on the bound state levels of the bubble in a MoS$_2$ monolayer. (a) Evolution versus bubble radius $R$. (b) Level structure and classification in terms of angular momentum and point of detachment from the unstrained bandstructure for a bubble of radius $R=6$ nm (shown in orange in a)). 
\label{Fig:BS-SOI}}
\end{center}
\end{figure}
 
\begin{figure}[t]
\begin{center}
\includegraphics[width=\columnwidth]{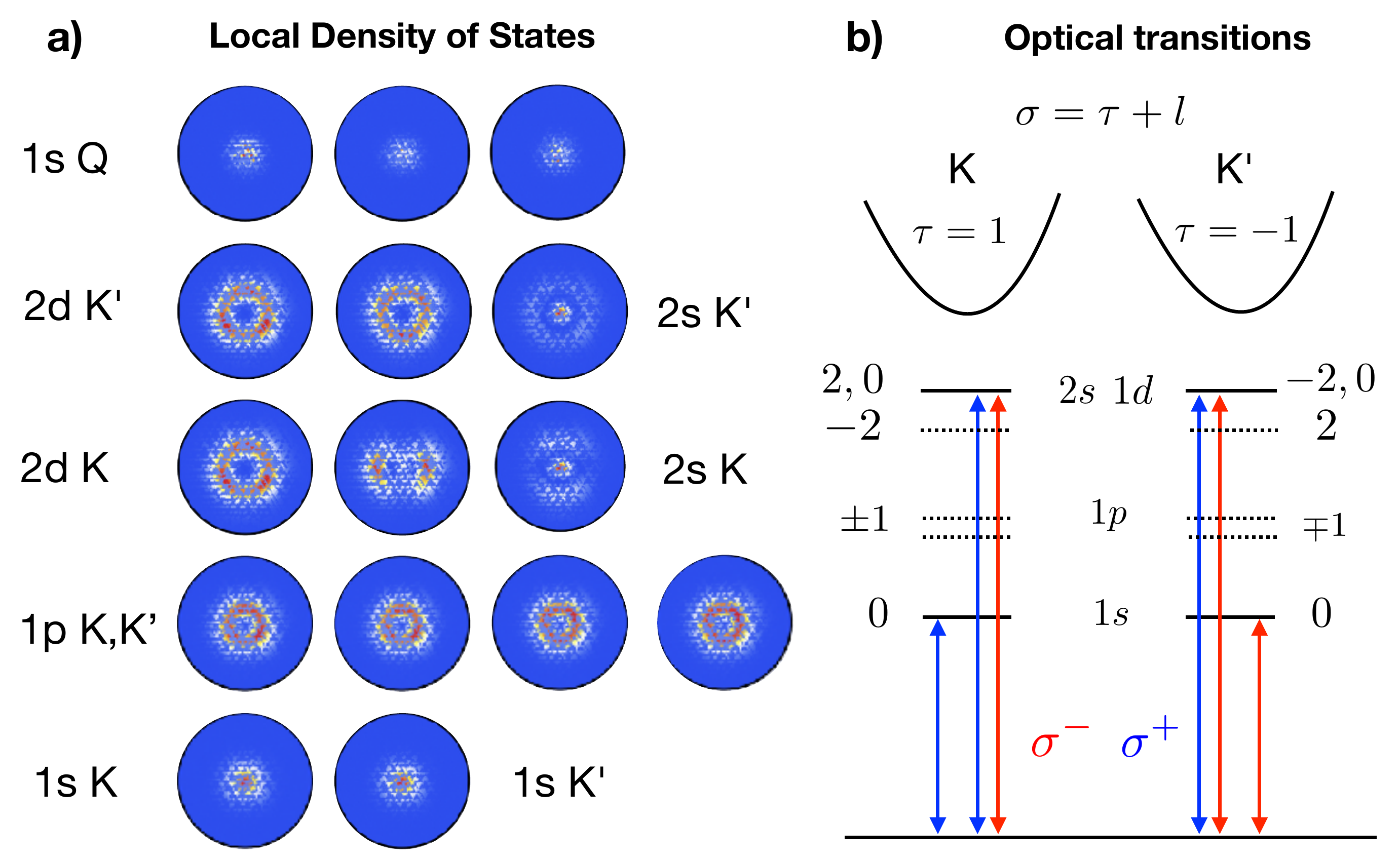}
\caption{Orbital structure of low energy bound states: (a) local density of states of the first 15 levels shown in Fig.~\ref{Fig:BS-SOI}b) for a radius of $R=6~{\rm nm}$. 
The size of the circle corresponds to $3R/2$. (b) Optical transition follow the rule $\sigma=\tau+l$, which gives rise to circular dichroism for the bubble states, similar to that of the unstrained material.  
\label{Fig:LDOSvsE}}
\end{center}
\end{figure}

The energy of K bound-states falls as $\epsilon(R)\approx \epsilon_0 + \alpha/R$ for some level-specific values of $\epsilon_0, \alpha$. They also follow a clear pattern in their degeneracy $d$. K-state energies are divided in groups of orbital degeneracy $d=g_v d_n$, with $g_v=2$ and $d_n=1,2,3,\ldots$ (without counting spin degeneracy for the moment). These are color coded in Fig.~\ref{Fig2} in orange ($d=2$), red ($d=4$), blue ($d=6$), etc.. The long wavelength nature of the strain perturbation guarantees that the valley index remains a good quantum number, providing a valley degeneracy $g_v=2$. This is confirmed also by the atomic orbital composition of the bound states, that show approximately a 90 \% Mo character and a 10 \% S character (not shown), compatible with states detaching from the $K$ and $K'$ valleys. By inspection of the spatial profile of the local density of states [with a maximum (node) at $R=0$ for the $s$-wave ($p$-wave, $d$-wave) states, see  Fig.~\ref{Fig:LDOSvsE} (a)], we establish that the $d=2$ states are valley degenerate $s$-wave solutions (zero angular momentum $l=0$), the $d=4$ group is composed by valley degenerate $p$-wave solutions with $l=\pm 1$, the $d=6$ group is composed by valley degenerate $s$-wave solutions and valley degenerate $d$-wave solutions with $l=\pm 2$.

The asymptotic behaviour of these states $\epsilon(R\to\infty)=\epsilon_0$ also supports the above picture. The effect of strain $\epsilon_\mathrm{max}$ at the center of the bubble is to locally reduce the minima of the conduction band, as shown in Fig. \ref{Fig1}(a-b). This reduction is expected to affect both the minima at the $K$ and $Q$ points. $s$-wave states that detach from the $K$ points saturate at large $R$ to a certain energy $\epsilon_0=E_K$ in the gap, while $s$-wave states that detach from the $Q$ points saturate to the higher $\epsilon_0=E_Q$ of $Q$ points under strain, see Fig. \ref{Fig1}. Note that the minimal $R\to\infty$ energy $E_K$ of confined states depends only on the aspect ratio of the bubbles [through $\epsilon_\mathrm{max}$, see Eq. \eqref{emax}], 
\begin{eqnarray}
E_K=E_K^0 - \gamma\frac{1}{2}\frac{\lambda+3\mu}{\lambda+2\mu}\left(\frac{h_{\rm max}}{R}\right)^2
\label{EK}
\end{eqnarray}
where $E_K^0$ is the conduction band edge for the TMD in question, and $\gamma$ is defined in Eq. \eqref{gamma}.

\begin{figure*}[t]
\begin{center}
\includegraphics[width=\textwidth]{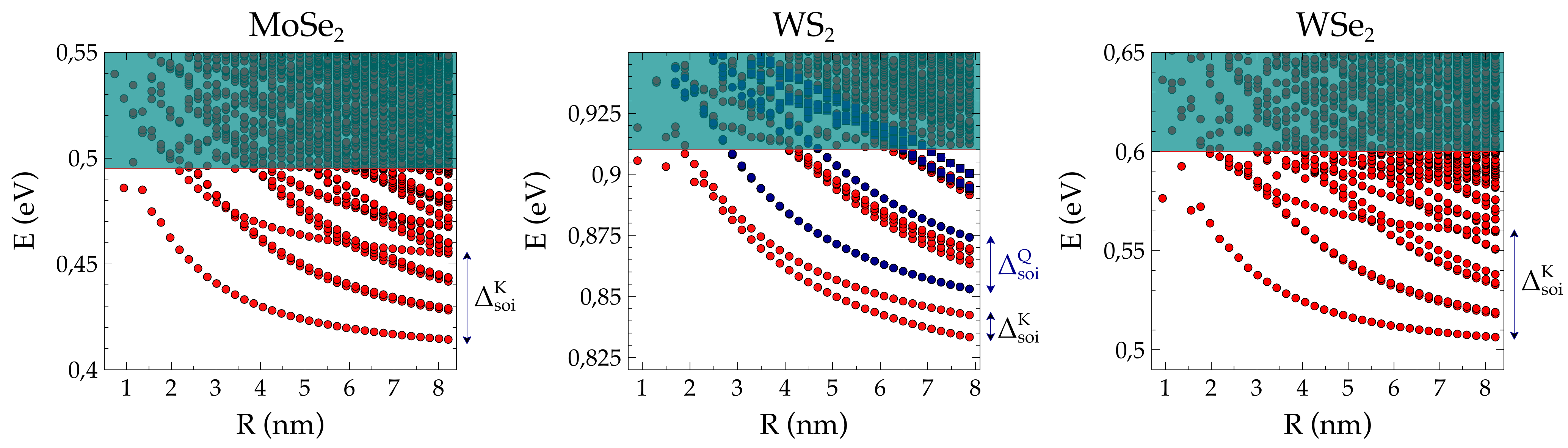}
\caption{Bound states evolution with bubble of aspect ratio $h_{\rm max}/R=0.14$ versus radius $R$ for monolayers of MoSe$_2$, WS$_2$, and WSe$_2$. For the case of the MoSe$_2$ and WSe$_2$ only states that detach from the $K$ and $K'$ points fall into the gap, whereas for the case of WS$_2$ also states that detach from the $Q$ points fall into the gap, with $s$-wave (circular marker) and $p$-wave (square marker) solutions. $\Delta^{K}_{\rm soi}$ and $\Delta^{Q}_{\rm soi}$ indicate the SOI gap that develops in the strained region at the $K$ and $Q$ points, respectively. \label{Fig:Panel-BS-SOI}}
\end{center}
\end{figure*}

{\it Classification of bound states.---} 
In the preceding section we computed the bubble electronic structure using a precise tight-binding model based on ab-initio results. Despite its complexity, we noted that the resulting level degeneracy and the level spacing of bubble bound states are compatible with a simple Fock-Darwin spectrum, characterised by a level degeneracy $d_n$ (save for a weak fine structure splitting), and alternating even and odd parity  $p_n=(-1)^{d_n+1}$ (circles and squares, respectively, in Fig. \ref{Fig2}). 

In this section we explain this structure using a simple effective-mass approximation for the bubble K/K' electrons, $H_\mathrm{eff}={\bf p}^2/2m^* + V(r)$. This model is parametrised by an effective mass $m^*$, and an effective parabolic confining potential $V(r)=m^* \omega_0^2r^2/2+\epsilon_0$, with $\omega_c$ the confining strength and $\epsilon_0=E_K$. The radius of the bubble defines the size of the confining potential $V(r)$, so that $\omega_0=\sqrt{2\epsilon_0/m^*}/R$.  The spectrum of $H_\mathrm{eff}$ is given by Fock-Darwin levels, $\epsilon_n=\hbar\omega_c(n+1)$, with $n=0,1,2,\ldots$ the principal quantum number. The cylindrical symmetry of the problem promotes a decomposition of the principal quantum number in $n=2n_r+|l|$, with the radial quantum number $n_r=0,1,2,\ldots$ and the azimuthal quantum number $l=-n_r,-n_r+2,\ldots,n_r-2,n_r$ [\onlinecite{Burkard_2018}]. It then follows that levels detach from the $K$, $K'$ points in the original Brillouin zone according to the law 
\begin{equation}\label{Eq:BSenergy}
\epsilon_{n_r,l}(R)=\frac{\hbar \sqrt{2\epsilon_0/m^*}}{R}(2n_r+|l|+1) + \epsilon_0,
\end{equation}
so that each principal quantum number $n$, characterized by a degeneracy $d_n=n+1$, provides a total $d=2d_n$ orbital degeneracy, when accounting for the valley degree of freedom for $K,K'$ points. The ground state is given by the combination $(n_r,l)=(0,0)$ and well describes the $d=2$ (orange) degenerate states that appear in the numerical simulations. Analogously, states given by the combination $(n_r,l)=(0,\pm 1)$ well agree with the $d=4$ (red)  group of levels. However, a closer inspection of Fig.~\ref{Fig2} shows a weak splitting between the $p^+$ and $p^-$ states. We ascribe this splitting to a Berry curvature effect, not included in $H_\mathrm{eff}$, that acts as an additional valley orbital angular momentum and separates $p^\pm$ states within the same valley (see Appendix). The $d=6$ group is formed by the valley degenerate two sets of states given by the combination $(n_r,l)=(0,\pm 2),(1,0)$, that according to Eq.~(\ref{Eq:BSenergy}) are expected to be degenerate. In fact, in the simulation the $s$-wave and $d$-wave solutions appear to admix and slightly split (see Fig.~\ref{Fig:LDOSvsE}). This is due to the fact that, within the same parity sector, long wavelength perturbations can split them.

{\it Spin-orbit interaction.---}
We now consider the effect of SOI, a very relevant perturbation in TMDs that substantially modifies the band structure. A good approximation is to neglect the spin-flipping terms, that couple low energy states to high energy bands due to parity change. It follows that $s_z$ is a good quantum number for the low energy states of the conduction and valence band\cite{Silva_AS_2016}. Since inversion symmetry is broken in TMDs, states at valleys $K$ and $K'$ are spin split in a time-reversal invariant way, so that spin $\uparrow$ states at valley $K$ are degenerate with spin $\downarrow$ states at valley $K'$, and viceversa. Analogously, SOI splits states at the $Q$ points, so that spin $\uparrow$ states at three $Q$ points surrounding valley $K$ along the 
three equivalent $\Gamma-K$ directions are degenerate with spin $\downarrow$ states at the three $Q'$ points surrounding valley $K'$, and viceversa. From an orbital point of view, this SOI mixes $d_{xy}$ and $d_{x^2-y^2}$ on the metal, and $p_x$, $p_y$ on the chalcogen sites, as analysed in Refs. \onlinecite{CG13,Silva_AS_2016}. 

In Fig.~\ref{Fig:BS-SOI}a we show the full subgap spectrum of a bubble on monolayer MoS$_2$ versus the radius $R$, including SOI. The spectrum remains time-reversal symmetric, so all levels remain at least two-fold Kramer's degenerate after adding SOI. The originally 4-fold  degenerate ground state (including spin) is split by SOI into Kramers doublets with opposite spins in opposite valleys, $|\uparrow K\rangle$ and $|\downarrow K'\rangle$, that could be used as spin-valley qubit \cite{Liu2014,YueWu2016,Szechenyi2018} -- an elastic analogue of quantum dot spin-qubits. The originally 8-fold first excited state (Fig.~\ref{Fig2}, red) is split by SOI in two groups of four states, that are then split by Berry curvature into Kramers doublets, and so on. 

Additionally, we see that, thanks to the SOI, states that detach from the $Q$ points can now bind strongly enough to enter the gap  (blue in Fig.~\ref{Fig:BS-SOI}). They are recognised by their offset energy from the K-states, their degeneracy (always a multiple of three), and by their orbital composition (approximately 80 \% Mo content and 20 \% S content, precisely the expected orbital contribution for the  Q points of the BZ.\cite{Silva_AS_2016}). We see that the six states ($|\uparrow Q_{1,2,3}\rangle$, $|\downarrow Q'_{1,2,3}\rangle$) are separated by $\Delta_{\rm soi}^Q$ from the six states ($|\downarrow Q_{1,2,3}\rangle$, $|\uparrow Q'_{1,2,3}\rangle$) by the strong SOI. This is due to the fact that the strong SOI lowers the energy of the SOI-split states at the $Q$ points relative to the K/K' points in the conduction band of the unstrained material, so that they can more easily fall into the gap by application of strain at the center of the bubble. 

In Fig.~\ref{Fig:BS-SOI}b we show the spectrum of a bubble of radius $R=6~$nm of the spin $\uparrow$ component and we specify the bound state angular momentum and the point of detachment in the unstrained band structure. The spectrum of the opposite spin component is identical with valley labels interchanged due to time-reversal symmetry. The spatial profile of the local density of state of the first 15 levels appearing in Fig.~\ref{Fig:BS-SOI}b is shown in  Fig.~\ref{Fig:LDOSvsE}a, where we recognise the different angular momentum solutions. 

In Fig.~\ref{Fig:Panel-BS-SOI} we finally show the spectrum versus bubble radius for the other three group VI-semiconducting TMDs, WSe$_2$, WS$_2$ and MoSe$_2$. We find a similar phenomenology, with levels that follow an overall Fock-Darwin spectrum, split by the strong SOI. For the case of MoSe$_2$ and WSe$_2$ only states that detach from the $K$ and $K'$ points fall into the gap, whereas for the case of WS$_2$ we observe that $s$-wave (circles) and $p$-wave (squares) Q-states fall into the gap.

{\it Optically selection rules.---} 
The bound states that detach from the valleys $K$ and $K'$ give rise to well defined selection rules for optical transitions. A valley angular momentum $\tau=\pm 1$ can be assigned to those states, so that optical transition are identical to those of the unperturbed material and satisfy $l+\tau=\sigma$, where $\sigma=\pm 1$ is the circular polarisation state of light. This rule is responsible for the  circular dichroism in these materials\cite{Mak2012,Zeng2012,Trushin_PRB_2016}. In particular,  $\sigma^\pm$ circularly polarised light couples to  $1s$ and $2s$ states in valley $\tau=\pm1$, and to $l=\pm 2$ $d$-wave states at valley $\tau=\mp 1$. Analogously, $p$-wave solutions are optically dark.

At the same time, states that detach from the minima at $Q$ have in-plane momentum that strongly differs from the top valence band states, that remain at $K$ even under strain. Therefore, only indirect transitions are allowed for $Q$ states, which renders them optical dark too. These considerations are important when comparing the spectral structure predicted in this work with photoemission and absorption measurements.

{\it Conclusions.---} 
We have analysed the elastic and electronic structure of circular bubbles formed spontaneously by gas or liquids trapped under monolayer TMD crystals. We have found that, as a result of the unique tendency of TMDs to reduce their bandgap under strain, bubbles are able to efficiently confine electronic states, which acquire an energy inside the gap and a rich symmetry structure. We find that their asymptotic binding energy $E_K^0-E_K$ depends quadratically on the aspect ratio of the bubbles, and the sensitivity $\gamma$ of the gap to strain, see Eq. \eqref{EK}. We have also identified a host of additional bound states of varying degeneracies and angular momentum components, with and without spin-orbit interaction. Particularly intriguing is a family of $Q$-point, optically dark states with a three-fold orbital degeneracy per spin component connected to the symmetry of the lattice under $120^\circ$ rotations. These states are strongly affected by spin-orbit interactions. We expect their degeneracy to be broken in non-circular bubbles, under uniaxial strain of the lattice, or by close proximity to other bubbles. These effects can serve as a very direct probe into the band structure of TMDs, as well as a versatile and realistic route towards electronic control through strain engineering. We further expect that the confinement of states found here for atomically thin bubbles will also be present in other structures such as 2D crystals deposited on substrates with nanodomes or nanopillars,\cite{Li_NC_2015,PB2016,Branny2016,Branny2017,Chaste2018,Tripathi2018,Burkard_2018} since the induced strain in the membranes is considerably higher than the one in naturally formed bubbles.

\section{Acknowledgments}

We acknowledge funding from the Graphene Flagship, contract CNECTICT-604391, from the Comunidad de Madrid through Grant MAD2D-CM, S2013/MIT-3007, from the Spanish Ministry of Economy and Competitiveness through Grants No. RYC-2011-09345, RYC-2016-20663, FIS2015-65706-P, FIS2016-80434-P (AEI/FEDER, EU) and the Mar\'ia de Maeztu Programme for Units of Excellence in R\&D (MDM-2014-0377).

\appendix

\section{Effect of Diracness}
\label{SubSec:Diracness}

In this section we provide an analysis beyond the parabolic conduction band approximation presented in the main text, that accounts for the valence band and the action of the strain as a local reduction of the semiconducting gap at the center of the bubble. Nevertheless, it is instructive to first consider a parabolic dispersion and a step-like potential well at the center of the bubble described by $\epsilon_c(r)=-\epsilon_c\theta(R-r)$. For $-\epsilon_c<\epsilon<0$ the solution for $r<R$ is a Bessel function of the first kind $J_l(kr)$, with $k=\sqrt{2m^*(\epsilon+\epsilon_c)}$ and for $r>R$ is a modified Bessel function of the second kind $K_l(qr)$, with $q=\sqrt{-2m^*\epsilon}$, that are well behaved at $r=0$ and $r\to \infty$, respectively. By matching the wavefunction and its derivative at $r=R$ we obtain the following implicit equation for the eigenvalues,
\begin{equation}\label{Bessl-p2}
qK'_{l}(qR)J_l(kR)-kJ'_l(kR)K_l(qR)=0.
\end{equation}  
The roots are given by the zeros of the Bessel function, that are characterised by the azimuthal quantum number $l=0,\pm1, \pm2,\ldots$ 
and a radial quantum number for each $l$, $n_l=0,1,2,\ldots$. It is clear from Eq.~(\ref{Bessl-p2}) that solution with same $|l|$ are strictly 
degenerate, whereas solutions belonging to the same parity sector are split by the step-like potential, that can be considered as a smooth perturbation 
on the atomic scale. In particular, it follows that Fock-Darwin levels $(2,0),(0,\pm 2)$ need not be degenerate, as it is observed in the simulations of Fig.~\ref{Fig2}. 

\begin{figure}[t]
\begin{center}
\includegraphics[width=0.45\textwidth]{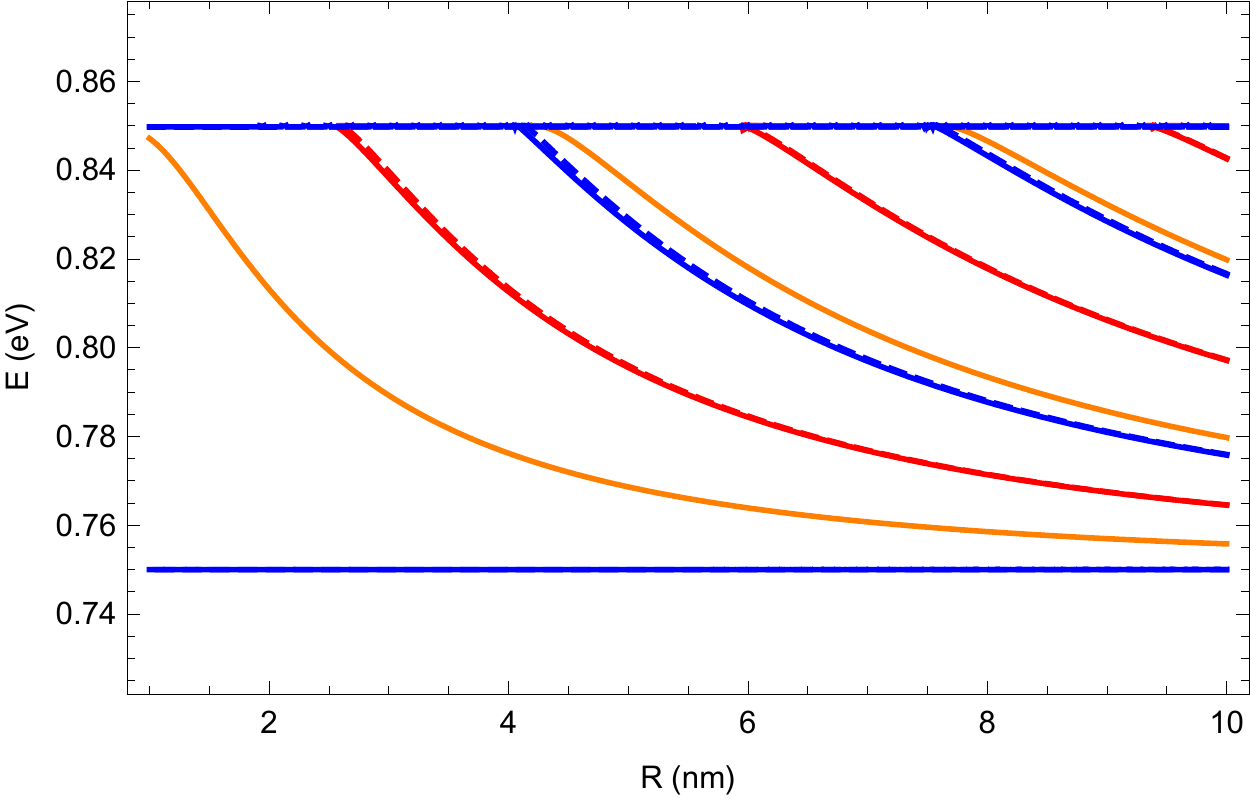}
\caption{Evolution of the bound state levels with the radius of the bubble $R$ for the gapped Dirac equation with position dependent gap, as given by the roots of Eq.~(\ref{RootsBessel}). Solutions with $l=0$ (orange), $\l=\pm 1$ (red solid/dashed), and $l=\pm 2$ (blue solid/dashed) are shown. A symmetric gap $\Delta=E^0_K=0.85~{\rm eV}$ has been chosen, the Dirac velocity has been set to $v=\sqrt{\Delta/m^*}$, with conduction band effective mass $m^*=0.49~m_e$, and the reduced gap value $\Delta_c\equiv E_K=0.75~{\rm eV}$ has been chosen at the bubble center. 
\label{Fig:BS_bessel}}
\end{center}
\end{figure}

In order to understand the splitting of the solution with same $|l|$ we need to go beyond the parabolic approximation and assume a two-band Dirac model for the valence and conduction band. The action of the strain is to modify the gap at the center of the bubble, which acquires a dependence on the position, $\Delta(r)$. For simplicity we now assume a step-like profile, $\Delta(r)=\Delta_c\theta(R-r)+\Delta\theta(r-R)$, with $R$ the radius of the bubble. The Hamiltonian for each valley reads
\begin{equation}
H_\tau =\left(\begin{array}{cc}
\Delta(r) & v(\tau p_x-ip_y)\\
v(\tau p_x+ip_y) & -\Delta(r)
\end{array}\right),
\end{equation}
with $\tau=\pm 1$ the valley index and $v$ the Dirac velocity. The problem is solved by matching the solutions for $r<R$ and $r>R$ at $r=R$. Writing $\tau p_x\pm ip_y=-ie^{\pm i \tau \phi}(\tau \partial_r\pm i\frac{1}{r}\partial_\phi)$, the eigenfunctions in the regions $r<R$ and $r>R$ are written in terms of Bessel functions $Z^{(j)}_l(x)$, 
\begin{equation}
\psi^{(j)}_{\tau,l}(r,\phi)=\frac{e^{il\phi}}{\sqrt{2\pi}}
\left(\begin{array}{c}
\alpha_{k,j} Z^{(j)}_l(kr)\\
\tau^j e^{i\tau\phi}\beta_{k,j} Z^{(j)}_{l+\tau}(kr)
\end{array}\right),
\end{equation}
where $j=0,1$  refer to $r<0$ and $r>R$, respectively, $Z^{(0)}_l(x)=J_l(x)$ and $Z^{(1)}_l(x)=K_l(x)$. The secular problem at energy $\Delta_c<\epsilon<\Delta$ is solved by $\alpha_{k,j}=\Delta_j+\epsilon$ and $\beta_{k,j}=-ivk_{j}$, with $\Delta_0=\Delta_c$, $\Delta_1=\Delta$, $vk_0=\sqrt{\epsilon^2-\Delta_c^2}$, and $vk_1=\sqrt{\Delta^2-\epsilon^2}$. Matching the wavefunctions at $r=R$ gives the following eigenvalues equation
\begin{equation}\label{RootsBessel}
\tau \frac{k_1(\Delta_c+\epsilon)}{k_0(\Delta+\epsilon)}=\frac{J_{l+\tau}(k_0R)K_l(k_1R)}{J_l(k_0R)K_{l+\tau}(k_1R)}.
\end{equation}
Similarity and differences with Eq.~(\ref{Bessl-p2}) are manifest, the confining action of the bulk evanescent solution is similar and the main difference stems from the mixing of the $l$ with $l\pm 1$ solutions, in a way that solutions with $\pm |l|$ are no longer degenerate within the same valley.

In Fig.~\ref{Fig:BS_bessel} we plot the roots of Eq.~(\ref{RootsBessel}) versus the radius $R$ for a single valley $\tau=1$ and for the values $l=0,\pm 1,\pm 2$. The spectrum captures very accurately the results of the simulations Fig.~\ref{Fig2}. The splitting between the $\pm|l|$ solutions is appreciated for small radius $R$ and the levels become degenerate in the limit of large $R$. The splitting at small $R$ is a consequence of the Diracness of the problem. At a given valley, the Berry curvature of the Dirac Hamiltonian provides an additional flux that treads with opposite sign states with opposite angular momentum. In the other valley the role of positive and negative angular momentum states is reversed, so as to preserve time-reversal symmetry, and the states $(\tau,l)$ and $(-\tau,-l)$ remain degenerate.

\bibliography{biblio.bib}

\end{document}